\documentclass[aps,prb,preprint,superscriptaddress,showpacs,showkeys,floatfix]{revtex4}

\usepackage{graphicx,color}

\newcommand{\rev}[1]{\textcolor{black}{#1}}

\graphicspath{{figs/}}
\begin{document}
\title{Strain Tunable Phononic Topological Bandgaps in Two-Dimensional Hexagonal Boron Nitride}
\author{Jin-Wu Jiang}
    \altaffiliation{Corresponding author: jwjiang5918@hotmail.com}
    \affiliation{Shanghai Institute of Applied Mathematics and Mechanics, Shanghai Key Laboratory of Mechanics in Energy Engineering, Shanghai University, Shanghai 200072, People's Republic of China}

\author{Harold S. Park}
    \altaffiliation{Corresponding author: parkhs@bu.edu}
    \affiliation{Department of Mechanical Engineering, Boston University, Boston, Massachusetts 02215, USA}

\date{\today}
\begin{abstract}

The field of topological mechanics has recently emerged due to the interest in robustly transporting various types of energy in a flaw and defect-insensitive fashion.  While there have been a significant number of studies based on discovering and proposing topological materials and structures, very few have focused on tuning the resulting topological bandgaps, which is critical because the bandgap frequency is fixed once the structure has been fabricated.  Here, we perform both lattice dynamical calculations and molecular dynamical simulations to investigate strain effects on the phononic topological bandgaps in two-dimensional monolayer hexagonal boron nitride.  Our studies demonstrate that while the topologically protected phononic bandgaps are not closed even for severely deformed hexagonal boron nitride, and are relatively insensitive to uniaxial tension and shear strains, the position of the frequency gap can be efficiently tuned in a wide range through the application of biaxial strains. \rev{Overall, this work thus demonstrates that topological phonons are robust against the effects of mechanical strain engineering,} and sheds light on the tunability of the topological bandgaps in nanomaterials.

\end{abstract}

\keywords{Mechanical Strain; Topology; Interface Phonon Mode; Hexagonal Boron Nitride}
\maketitle
\pagebreak

\section{Introduction}

Topological insulators (TIs) represent a new state of matter whose behavior depends only on its topology, rather than its geometry~\cite{hazanRMP2010,mooreNATURE2010,qiPT2010,wangNM2017}.  Since their theoretical prediction and subsequent experimental realization a little more than a decade ago~\cite{kanePRL2005a,bernevigSCIENCE2006}, TIs have attracted significant interest due to their unique and exciting physical properties, namely being an insulator in the bulk while simultaneously enabling wave propagation along its boundary~\cite{mooreNATURE2010}.  Because these surface states on the boundary of TIs are topologically protected, they are robust in the presence of defects, and have generated significant interest in controlling and guiding the propagation of acoustic, electronic, and phononic waves that are immune not only to backscattering, but also to structural defects like cavities, disorder and sharp bends~\cite{heNP2016,qiPT2010}.   

Because of these unique properties, in the past five years there has been an explosion of work that has focused on making connections between the quantum mechanical and classical pictures of transport~\cite{mousaviNC2015,huberNP2016,susstrunkPNAS2016,heNP2016}.  In particular, researchers have 
used principles underlying the quantum hall effect~\cite{klitzingPRL1980,wangPRL2015,nashPNAS2015}, the quantum spin hall effect~\cite{haldanePRL1988,kanePRL2005a,bernevigSCIENCE2006,mousaviNC2015,susstrunkPNAS2016,susstrunkSCIENCE2015,palJAP2016,palARXIV2017,yuARXIV2017,prodanNC2017,husseinAMR2014,heNP2016,cummerNRM2016,palNJP2017,xiaoNP2015}, and the quantum valley hall effect~\cite{renRPP2016,palNJP2017,liuARXIV2017,wuARXIV2017} to generate topologically protected phonons in mechanical structures and metamaterials.  While nearly all of these works have been at the macroscale, the present authors have recently demonstrated the existence of topologically protected phonons in the two-dimensional nanomaterial hexagonal boron nitride (h-BN).\cite{JiangJW2017toph-bn}

However, to-date, most works have focused on designing or discovering topologically protected phonons in different mechanical structures or metamaterials.  In contrast, relatively few works, with notable exceptions~\cite{susstrunkNJP2017}, have investigated the tunability of the resulting topological bandgaps.  This functionality is important for practical applications because the topological properties, and specifically the frequency range of the topological bandgap, are fixed once a specific mechanical structure has been fabricated.  

Therefore, in this work, we examine, using both molecular dynamics and lattice dynamics calculations, the effects of mechanical strain on the topological phonon modes in two-dimensional hexagonal boron nitride.  Examining the effects of strain is considerably simpler in this two-dimensional material because, unlike many macroscale realizations of phononic topological insulators which are based on discrete elements that are not physically connected, the boron and nitrogen atoms in h-BN are connected by covalent bonds whose strength depends on the strain that is applied to the atomic crystal.  We study the effect of mechanical strain on the topological phonon modes that cross over the frequency gap [1123, 1278]~{cm$^{-1}$} in the phonon spectrum, which are localized at the topological interface for the h-BN. We find that the frequency gap is robust against biaxial, uniaxial, and shear strains, i.e. the frequency gap is not closed by any of these mechanical deformations up to fairly large strains.  Thus, we demonstrate that while the topologically protected phonon modes are stable, they can simultaneously be efficiently tuned by biaxial strain, which reduces the upper bound frequency of the topological bandgap from 1200~{cm$^{-1}$} to 700~{cm$^{-1}$} for 12\% biaxial strain.

\section{Lattice dynamics calculations}

\begin{figure*}[htpb]
  \begin{center}
    \scalebox{1.0}[1.0]{\includegraphics[width=\textwidth]{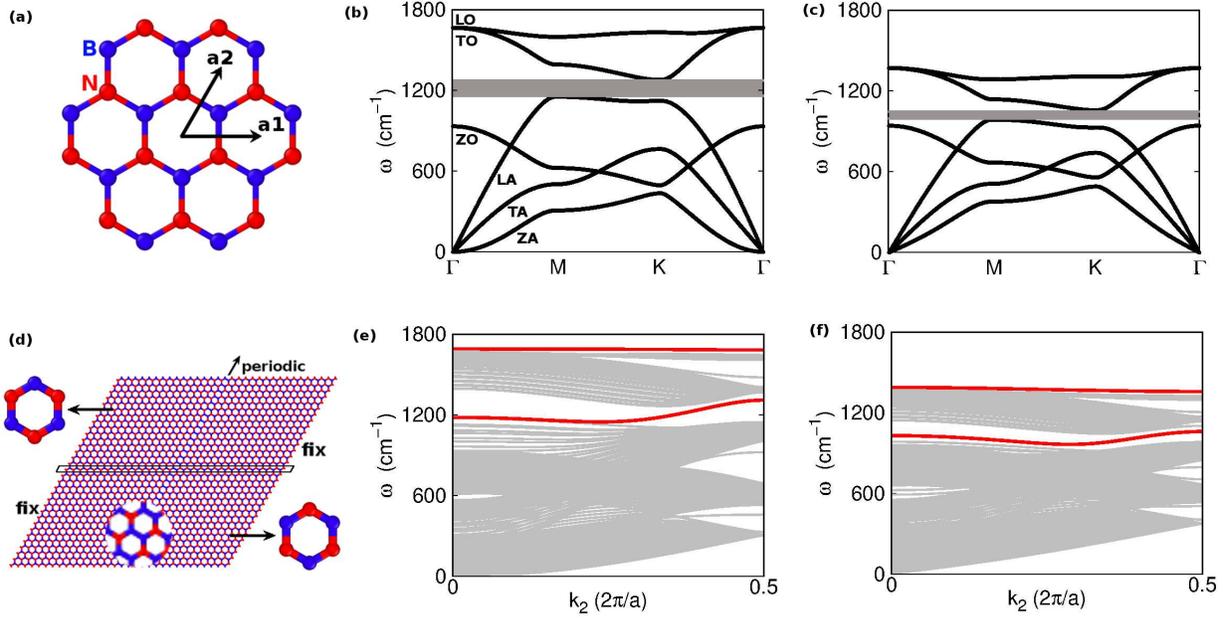}}
  \end{center}
  \caption{(Color online) Strain effects on the phonon dispersion of h-BN. (a) The hexagonal lattice structure of h-BN, with basis vectors $\vec{a}_1$ and $\vec{a}_2$. (b) Phonon dispersion for the undeformed h-BN. The frequency gap [1123, 1278]~{cm$^{-1}$} is illustrated by the grey area. (c) Phonon dispersion for h-BN deformed by a biaxial strain of $\epsilon=0.05$. The frequency gap is shifted downwards and narrowed. (d) A h-BN ribbon with topological B-B interface (enlarged by the bottom circular inset) along the $\vec{a}_2$ direction.  The h-BN lattices on the left and right sides of the topological interface are different due to an inversion operation.  The unit cell of the ribbon is enclosed by a parallelgram. (e) Phonon dispersion for the h-BN ribbon shown in (d) without deformation. Phonon branches at the interface are highlighted by red thick lines. Note the topologically protected interface modes crossing over the frequency gap [1123, 1278]~{cm$^{-1}$}. (f) Phonon dispersion for the h-BN ribbon deformed by the biaxial strain $\epsilon=0.05$, with topologically protected interface modes crossing over the frequency gap [987, 1057]~{cm$^{-1}$}.}
  \label{fig_phonon_strain}
\end{figure*}

The h-BN has a honeycomb lattice structure of D$_{\rm 3h}$ symmetry as shown in Fig.~\ref{fig_phonon_strain}~(a). The primitive unit cell is denoted by the two basis vectors $\vec{a}_1=a\hat{e}_x$ and $\vec{a}_2=a(\frac{1}{2}\hat{e}_x + \frac{\sqrt{3}}{2}\hat{e}_y)$, with $a=2.50$~{\AA} as the lattice constant. The x-axis is along the horizontal direction, while the y-axis is along the vertical direction. The two atoms (B and N) in the primitive unit cell are different, so inversion symmetry is broken for h-BN.

Fig.~\ref{fig_phonon_strain}~(b) shows the phonon dispersion for h-BN, where the atomic interactions for h-BN are described by the Tersoff potential.\cite{LindsayL2011prb} We note the opening of the frequency gaps for the two Dirac-like dispersions at the K point, which emerge due to the broken inversion symmetry for the primitive unit cell of h-BN.  The higher frequency gap [1123, 1278]~{cm$^{-1}$} (grey area) is of particular importance, because there is no other phonon branch falling within this frequency gap. Fig.~\ref{fig_phonon_strain}~(c) shows the phonon dispersion of h-BN that has been stretched biaxially with strain $\epsilon=0.05$. There are two distinct features in the phonon dispersion of the deformed h-BN. First, the frequency gap is robust against the biaxial tension, although the gap is obviously narrowed by the strain engineering. Second, the position (frequency) of the gap can be effectively tuned by the biaxial strain, i.e., the frequency gap shifts downwards due to the biaxial tension.

Fig.~\ref{fig_phonon_strain}~(d) shows a monolayer h-BN sheet denoted by $n_1 \vec{a}_1 \times n_2 \vec{a}_2$. The structure shown in the figure has the size $32 \vec{a}_1 \times 32 \vec{a}_2$. The left and right ends are fixed for the lattice dynamical calculation, while periodic boundary conditions are applied along the $\vec{a}_2$ direction. The big unit cell enclosed by the black box is used for the  phonon calculation. There is an interface along the $\vec{a}_2$ direction in the middle of the structure, which divides the structure into the left and right regions with different topology (see left top and right bottom insets). This type of interface will be referred to as the topological interface.  More specifically, we have previously shown that the Berry curvature for the lower boundary phonons of the frequency gap for h-NB (i.e., B and N are switched as compared to h-BN as in the right bottom inset of Fig.~\ref{fig_phonon_strain}~(d)) have opposite sign as compared to h-BN on the left top inset of Fig.~\ref{fig_phonon_strain}~(d).  Thus, topologically protected localized modes exist along the interface of the h-BN and h-NB lattices with different valley Chern numbers, which is in analogy with the quantum valley hall effect.\cite{JiangJW2017toph-bn}

Fig.~\ref{fig_phonon_strain}~(e) displays the phonon dispersion for the topological interface shown in Fig.~\ref{fig_phonon_strain}~(d). There are 192 branches corresponding to the 64 atoms in the big unit cell, and the red lines depict the phonon modes that are localized at the interface. There is a particular interface phonon branch crossing over the frequency gap [1123, 1278]~{cm$^{-1}$}, which are the topologically protected interface phonons.\cite{JiangJW2017toph-bn} \rev{We note that this frequency is in the optical range. The interface has low symmetry, so the topological phonon is Raman active and may thus be detected experimentally by Raman scattering.}

\begin{figure}[htpb]
  \begin{center}
    \scalebox{1.0}[1.0]{\includegraphics[width=8.5cm]{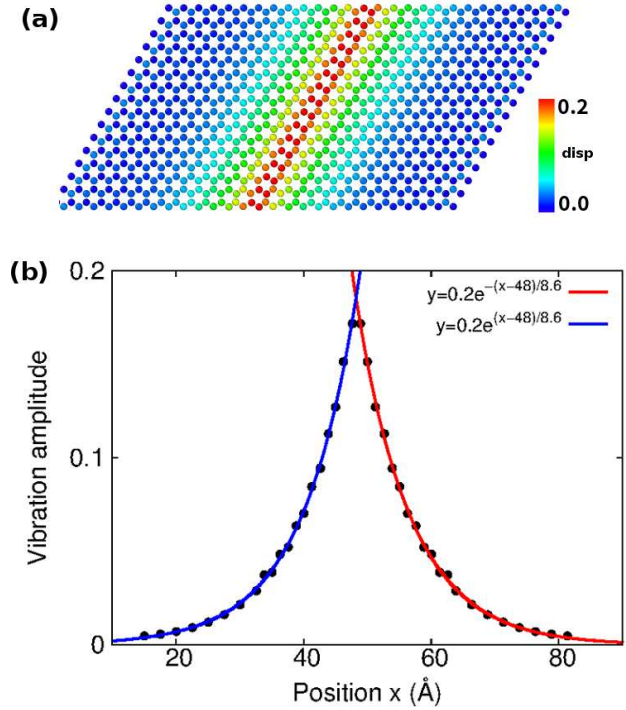}}
  \end{center}
  \caption{(Color online) Vibrational morphology for the topological interface mode with frequency 1182~{cm$^{-1}$} at $k_2=0$ point. (a) The eigenvector of the interface mode at $k_2=0$. The color bar represents the value of the eigenvector. (b) An exponential decay (from the interface) for the amplitude of the eigenvector in (a). The exponential factor is 8.6~{\AA}, which can be regarded as the localization length of the localized mode.}
  \label{fig_vibration_morphology}
\end{figure}

\begin{figure*}[htpb]
  \begin{center}
    \scalebox{1.0}[1.0]{\includegraphics[width=\textwidth]{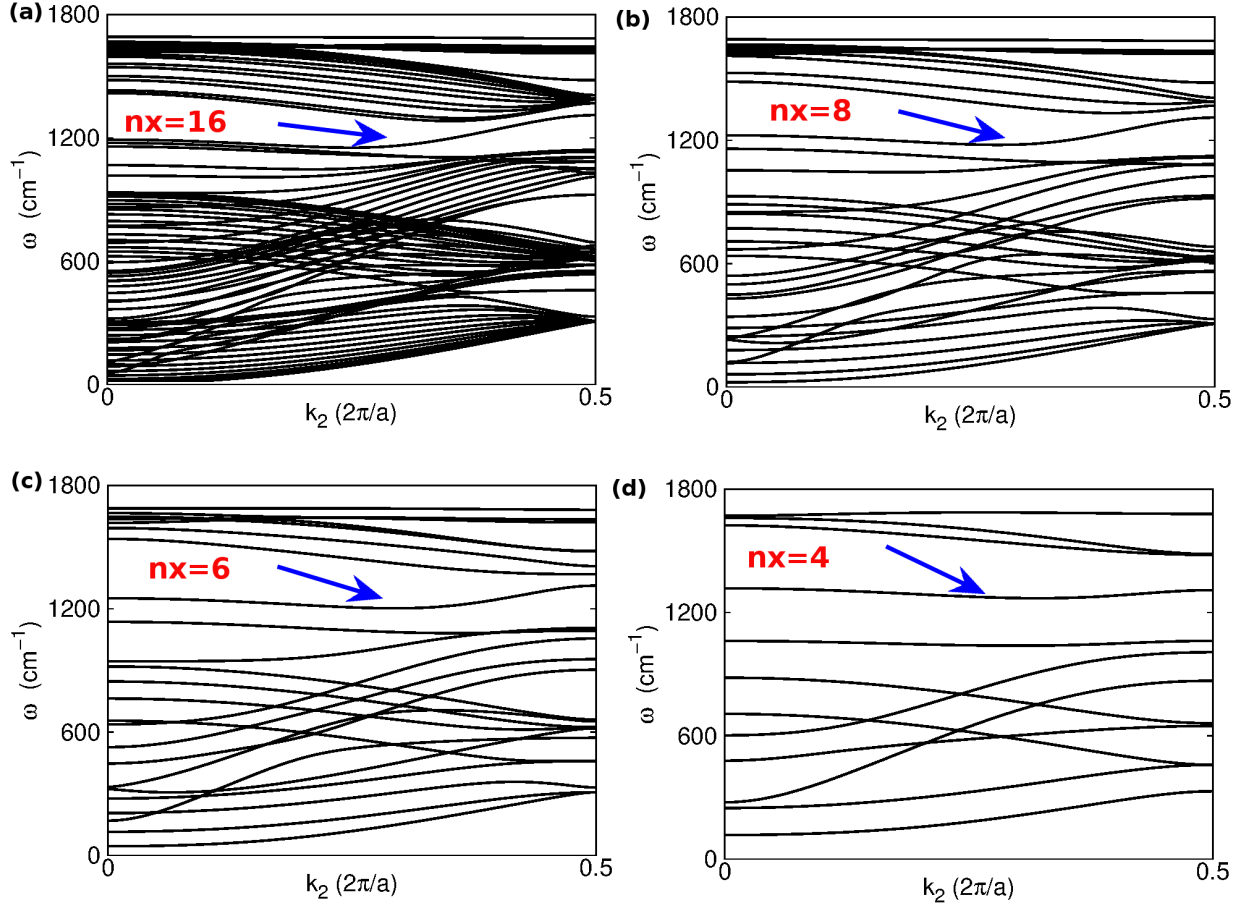}}
  \end{center}
  \caption{(Color online) The effect of the super cell size ($n_x$) on the topological interface branch. From (a) to (d): the number of the unit cell is $n_x=$ 16, 8, 6, and 4.}
  \label{fig_nx_effect}
\end{figure*}

\rev{The topological interface phonon is localized at the interface. Fig.~\ref{fig_vibration_morphology} shows that the eigenvector of the topological interface mode decays exponentially away from the interface, which is a characteristic feature for the localized mode.\cite{BornM} In particular, the localization length of this localized mode is $l_c=8.6$~{\AA}, so we have $l_c/a \approx 3$ with $a=2.50$~{\AA} as the lattice constant of the bulk h-BN. The localized nature of the topological interface mode implies that this mode will not be affected by the size ($n_x$) of the super cell if $n_x$ is large enough compared with the localization length $l_c$. However, if the size ($n_x$) of the super cell is too small (i.e., comparable or less than $3\times 2 = 6$), then $n_x$ will impact the interface mode. Fig.~\ref{fig_nx_effect} compares the phonon dispersion of the h-BN with topological B-B interface at the middle of the super cell $n_x\times 1$ with $n_x=$ 16, 8, 6, 4. It shows that the topological interface phonon branch crosses over the frequency gap perfectly for $n_x=16$, which is almost the same as the results for $n_x=32$. The size $n_x$ starts to take effect for $n_x=$ 8 and 6. For $n_x=4$, the interface phonon is strongly impacted by the size effect, as the size of the super cell is smaller than the localization length ($l_c$) of the interface phonon. In practice, the size of the system is much larger than the localization length of the interface phonon, so large values of $n_x$ are typically used so that it has no effect on the topological interface mode. We thus use $n_x=32$ in actual calculations in this work.}

Fig.~\ref{fig_phonon_strain}~(f) shows the phonon dispersion for the h-BN sheet that is biaxially stretched for $\epsilon=0.05$. The frequency gap [987, 1057]~{cm$^{-1}$} is narrowed and shifted downwards by the biaxial tension. Interestingly, the topologically protected phonon branch is robust and still crosses over the frequency gap in the deformed h-BN sheet.

\begin{figure*}[htpb]
  \begin{center}
    \scalebox{1.0}[1.0]{\includegraphics[width=\textwidth]{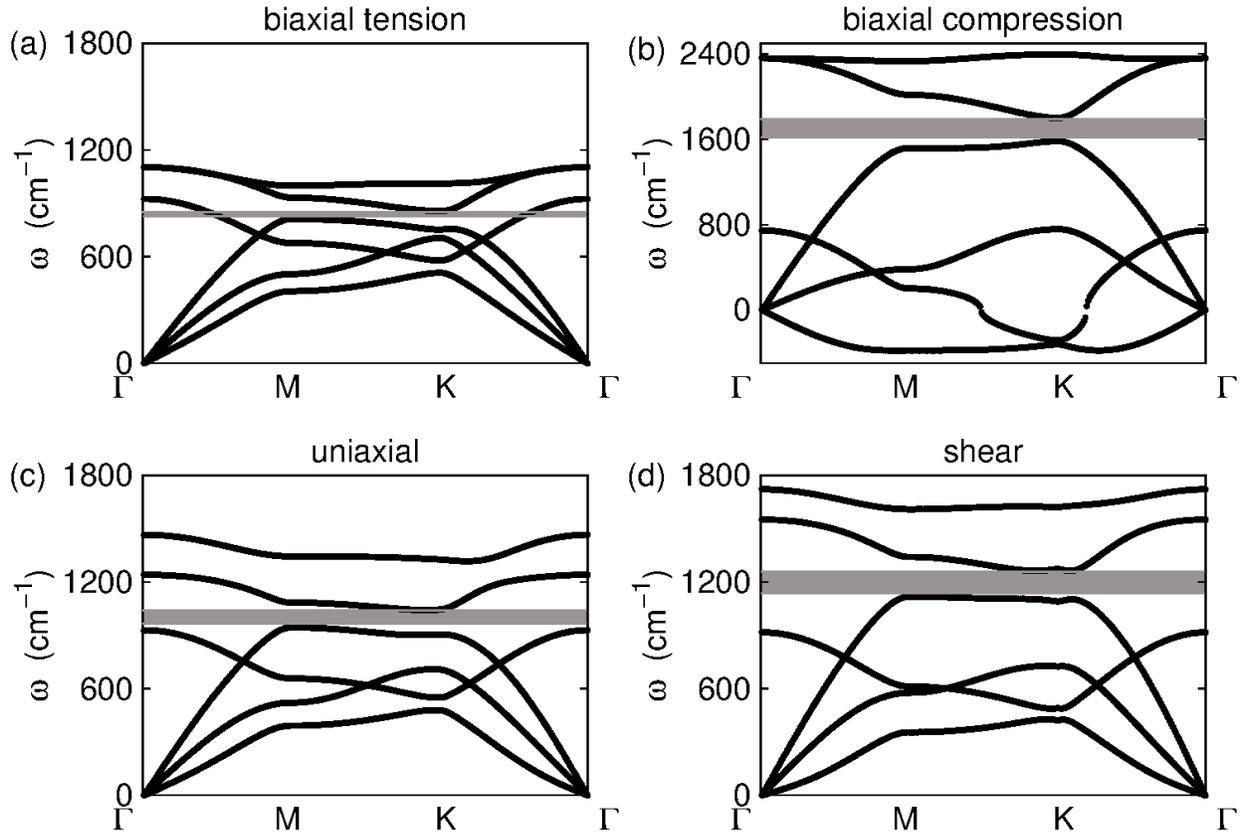}}
  \end{center}
  \caption{(Color online) The robustness of the topological bandgap (grey area) in h-BN against different types of deformation: (a) biaxial tension, (b) biaxial compression, (c) uniaxial tension, and (d) shear strain. The magnitude of the strain is 0.1 for all deformations.}
  \label{fig_phonon_deformation}
\end{figure*}

To further demonstrate the robustness of the frequency gap against various types of mechanical deformations, we study in Fig.~\ref{fig_phonon_deformation} the phonon dispersion of h-BN that is deformed by biaxial tension, biaxial compression, uniaxial strain, and shear strain. The magnitude of the strain is 0.1 for all deformations. \rev{The deformed structure is optimized to the energy minimum state, based on which the phonon dispersion is calculated.} Fig.~\ref{fig_phonon_deformation} shows that the frequency gap is rather robust against all of these deformations, and that the frequency gap is not closed, though it becomes incomplete, which we will discuss next, despite the large magnitude of the applied strain ($\epsilon=0.1$).  

\rev{There are six branches, including the z-directional acoustic (ZA), the transverse acoustic (TA), the longitudinal acoustic (LA), the z-directional optical (ZO), the transverse optical (TO), and the longitudinal optical (LO) branches.} We note that the frequency gap in Fig.~\ref{fig_phonon_deformation}(a) is between two in-plane vibrational branches (TO and LA).  The ZO branch emerges within the frequency gap in Fig.~\ref{fig_phonon_deformation}(a), but this ZO branch does not affect the topological nature of the frequency gap.  We also note the presence of some imaginary phonon modes in the out-of-plane direction in Fig.~\ref{fig_phonon_deformation}~(b), which implies the instability (buckling along the out-of-plane direction) of the h-BN under compression. However, we will illustrate later that the robustness of the topologically-protected phonon modes does not depend upon the existence of a complete frequency gap.  The origin of the frequency gap is the different mass of the B and N atoms in the unit cell of the h-BN. As a result, while the frequency gap is incomplete due to the applied biaxial strain, it is robust and cannot be closed through strain engineering. This finding is similar to other works showing the robustness of the valley-dependent topologically protected phonons despite the incomplete bandgap.~\cite{liuARXIV2017} \rev{It is because the in-plane topological phonons can be excited while the ZO mode is not affected, as the coupling between the in-plane mode and the out-of-plane mode is weak.}  According to Fig.~\ref{fig_phonon_deformation}, the biaxial tension is the most effective approach to tune the frequency position and magnitude of the topological bandgap, and thus we focus on investigating the effects of biaxial tension on the topologically protected phonons in the h-BN. \rev{We note that there is a new bandgap between the TO and LO branches near the $\Gamma$ point in Fig.~\ref{fig_phonon_deformation}~(c) and (d). It is because the degeneracy of the TO and LO modes is due to the three-rotational symmetry of the h-BN lattice structure, which is broken by the uniaxial or shear strain. As a result, the degeneracy between the TO and LO branches is removed, leading to the new bandgap.}

\begin{figure*}[htpb]
  \begin{center}
    \scalebox{1}[1]{\includegraphics[width=\textwidth]{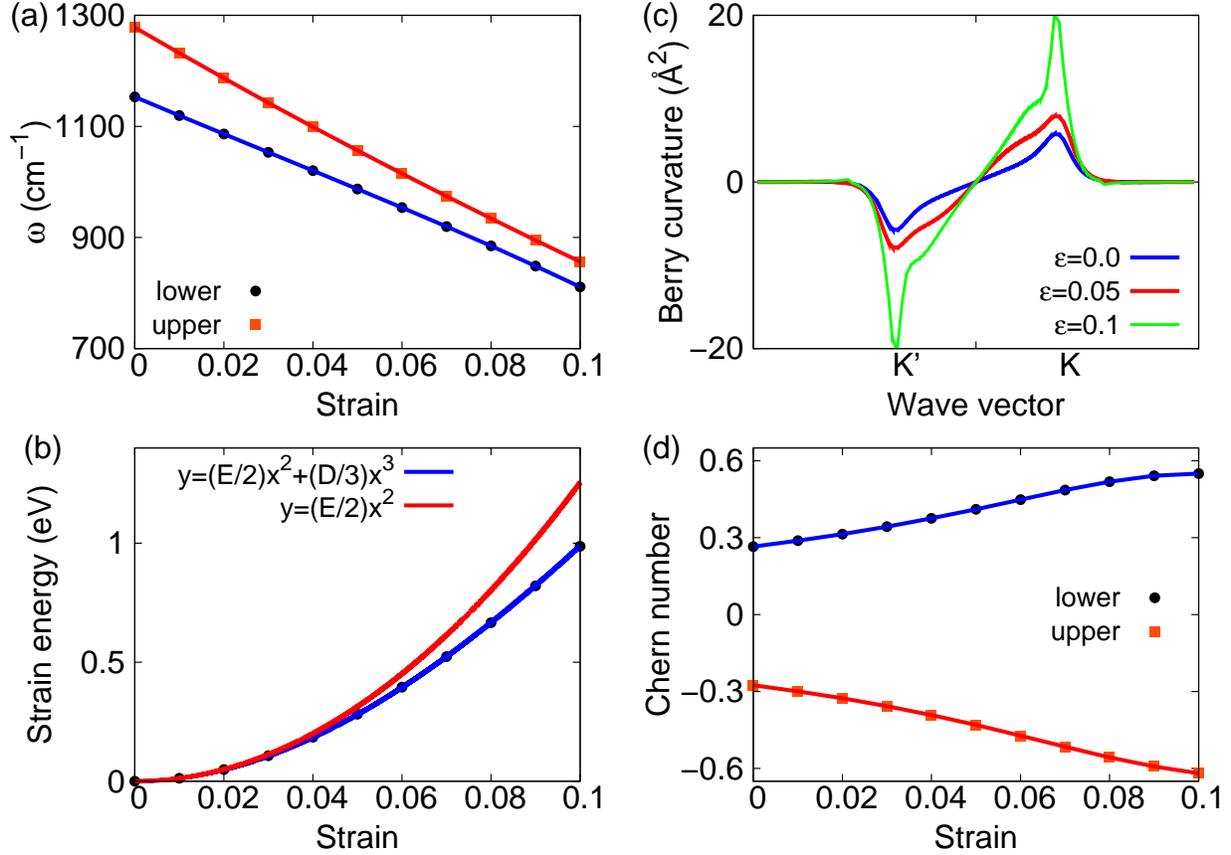}}
  \end{center}
  \caption{(Color online) Strain dependence of the topological properties for biaxially strained h-BN. (a) Strain dependence for the lower and upper boundaries of the topological bandgap of h-BN, where the upper bound corresponds to the TO branch and the lower bound corresponds to the LA branch in Fig.~\ref{fig_phonon_deformation}(a). (b) The strain energy (per unit cell) for the biaxially stretched h-BN. A parabolic fitting (red online) shows obvious deviation from the cubic fitting (blue online), showing the importance of the cubic nonlinear effect. Fitting parameters are $E=251$~{eV} and $D=-796$~{eV}. (c) The Berry curvature for the lower phonon branch of the bandgap in the biaxially stretched h-BN. (d) Strain dependence for the Chern number of the lower and upper boundaries of the topological bandgap of h-BN.}
  \label{fig_quantity_strain}
\end{figure*}

Fig.~\ref{fig_quantity_strain}~(a) shows the biaxial strain dependence for the frequency of the lower and upper boundaries of the topological bandgap in h-BN. The lower boundary is tuned gradually from 1153~{cm$^{-1}$} to 733~{cm$^{-1}$}, while the upper boundary shifts from 1278~{cm$^{-1}$} to 781~{cm$^{-1}$}. The frequency gap is narrowed gradually by the biaxial strain, and the magnitude of the frequency gap decreases from 125~{cm$^{-1}$} to 48~{cm$^{-1}$} from 0 to 12\% strain.  We note that even for extremely large biaxial strains like $\epsilon=0.2$, the frequency gap remains open, with the gap being 35~{cm$^{-1}$}.

The mechanical tunability of the frequency as well as the frequency gap is achieved through the cubic nonlinear interaction in the h-BN. Fig.~\ref{fig_quantity_strain}~(b) shows the strain energy per unit cell in h-BN, in which the cubic nonlinear effect plays an important role. The strain energy can be accurately fitted to the function $V(\epsilon)=\frac{1}{2}E\epsilon^2+\frac{1}{3}D\epsilon^3$, with parameters $E=251$~{eV} and $D=-796$~{eV}. The cubic nonlinear term can be absorbed by renormalizing the Young's modulus, i.e., $E'=E(1+\frac{2D}{3E}\epsilon)$. As a result, the effect of the cubic nonlinearity is to reduce the value of the Young's modulus, which results in the red-shift of the frequency gap in the h-BN. More generally, a larger contribution of the nonlinear term to the strain energy would result in a larger shift in the frequency of the topological bandgap.

Fig.~\ref{fig_quantity_strain}~(c) shows the Berry curvature along the KK$'$ line in the Brillouin zone for the lower phonon branch (i.e., the LA branch in Fig.~\ref{fig_phonon_deformation}(a)) of the bandgap in the h-BN biaxially stretched by strain $\epsilon=$ 0.0, 0.05, and 0.1. \rev{The Berry curvature for the phonon mode indexed by $\tau$ at the wave vector $\vec{k}$ is calculated by\cite{PalRK2017njp}
\begin{eqnarray}
B^{\tau}\left(\vec{k}\right) = -2Im\sum_{\tau'\not=\tau}\frac{\langle\tau|\frac{\partial D}{\partial k_{x}}|\tau'\rangle\langle\tau'|\frac{\partial D}{\partial k_{y}}|\tau\rangle}{\left(\omega_{\tau}^{2}-\omega_{\tau'}^{2}\right)^{2}},
\label{eq_berry}
\end{eqnarray}
where $D$ is the dynamical matrix. $\omega_{\tau}$ and $|\tau\rangle$ are the frequency and the polarization vector of the phonon mode $\tau$, respectively. The valley Chern number is computed by integrating the Berry curvature over a small region near the K and K$'$ points as,
\begin{eqnarray}
C_{\nu}^{\tau} & = & \frac{1}{2\pi}\int_{\nu}B^{\tau}\left(\vec{k}\right)d\vec{k},
\label{eq_chern}
\end{eqnarray}
where $\nu=$ K, K$'$ is the valley index.}

We find that the Berry curvature is increased and becomes more localized around the K and K$'$ points with the increase of the biaxial strain. The resultant valley Chern number at K point increases with increasing biaxial strain as shown in Fig.~\ref{fig_quantity_strain}~(d). The valley Chern number is obtained by integrating in the $\vec{k}$ space over a circular area with $|\vec{k}-\vec{K}| < |\vec{KK'}|/2$ near the K point. For $\epsilon=0$, the valley Chern number deviates from $\pm 0.5$, because the Berry curvature is not ideally localized around the K or K$'$ points as shown in Fig.~\ref{fig_quantity_strain}~(c).

\section{Molecular dynamics simulations}
From the above discussion, we have observed the topological phonon branch crossing over the frequency gap [1123, 1278]~{cm$^{-1}$} for the topological interface in unstrained h-BN. We have also demonstrated the robustness and the tunability of the topological phonon branch through strain engineering. To verify the lattice dynamical calculations, we perform classical molecular dynamics (MD) simulations to study the energy transfer along the topological interface in h-BN. The standard Newton equations of motion are integrated in time using the velocity Verlet algorithm with a time step of 1~{fs}. Simulations are performed using the publicly available simulation code LAMMPS~\cite{PlimptonSJ}, while the OVITO package is used for visualization~\cite{ovito}.  The interatomic interactions for h-BN are again modeled using the Tersoff potential.\cite{LindsayL2011prb}

\begin{figure}[htpb]
  \begin{center}
    \scalebox{1.3}[1.3]{\includegraphics[width=8cm]{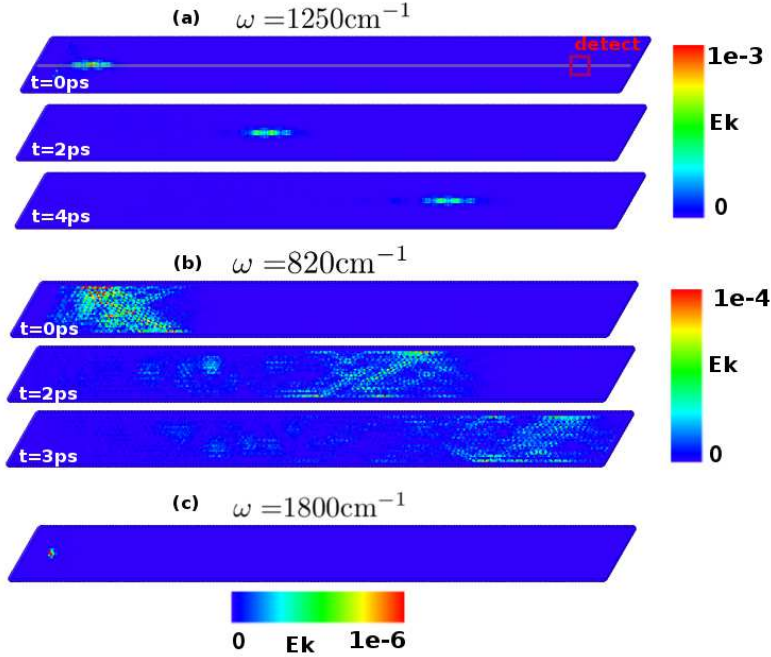}}
  \end{center}
  \caption{(Color online) MD snapshots for the vibrational energy transport along the topologically protected interface in the undeformed h-BN ribbon. The vibrational energy is excited at three representative frequencies, including (a) $\omega=1250$~{cm$^{-1}$} (within the topologically protected frequency gap), (b) $\omega=820$~{cm$^{-1}$}, and (c) $\omega=1800$~{cm$^{-1}$}. }
  \label{fig_md_strain_0.0}
\end{figure}

To examine strain effects on the topologically protected transport of phonons in biaxially strained h-BN, we created a h-BN sheet of size $200 \vec{a}_1 \times 20 \vec{a}_2$ as shown in Fig.~\ref{fig_md_strain_0.0}~(a).  The topological interface is along the horizontal direction in the middle of the h-BN sheet.  \rev{Waves with specified frequency were generated at the left end of the interface by driving one atom to oscillate at the given frequency for 30 cycles, which is modulated by the Hanning window. The vibrational energy transported is detected at the right end of the interface (area enclosed by a square), from which the transmission is computed.}

\begin{figure}[htpb]
  \begin{center}
    \scalebox{1}[1]{\includegraphics[width=8cm]{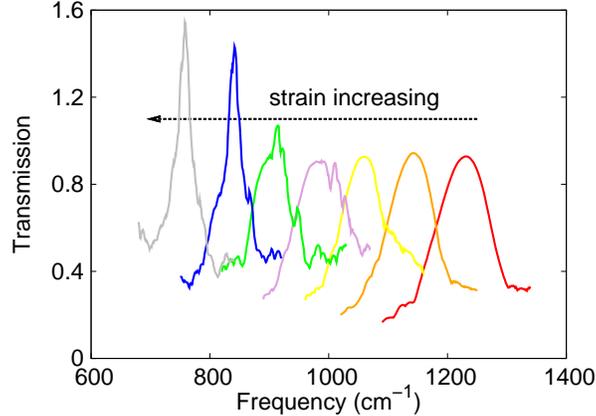}}
  \end{center}
  \caption{(Color online) Strain effect on the transmission function. From right to left, the applied biaxial strains are 0.0, 0.02, 0.04, 0.06, 0.08, 0.1, and 0.12.}
  \label{fig_transmission}
\end{figure}

The transmission is shown in Fig.~\ref{fig_transmission} for biaxially strained h-BN.  By comparing the transmission, it is clear that the working frequency of the topologically protected interface phonons can be manipulated via biaxial strain. Furthermore, these topological modes are rather robust and are still able to transport vibrational energy even for large biaxial strains in h-BN. There are some fluctuations in the transmission for large strains, and the transmissions for very high strains (i.e., over 0.1) can be larger than 1.0. This is because the large biaxial tension increases the vibrational frequency of the atoms, and thus the measured kinetic energy.

Some MD snapshots are presented in Fig.~\ref{fig_md_strain_0.0} for the energy transfer along the topological interface of the undeformed h-BN. In Fig.~\ref{fig_md_strain_0.0}~(a), the vibrational energy excited at frequency $\omega=1250$~{cm$^{-1}$} (which is within the topologically protected frequency gap) travels along the interface from left to right, with a high rate of transmission of 92\% as shown in Fig.~\ref{fig_transmission}.  In Fig.~\ref{fig_md_strain_0.0}~(b), the vibrational energy excited at frequency $\omega=820$~{cm$^{-1}$} is a normal phonon mode, which falls outside of the topologically protected frequency range. This normal mode is spatially extended, so the transmission at $\omega=820$~{cm$^{-1}$} is smaller than the topologically protected phonons, with a transmission of 15\%. Fig.~\ref{fig_md_strain_0.0}~(c) shows that it is rather difficult to inject energy into the h-BN by vibrating at frequency $\omega=1800$~{cm$^{-1}$}, which is outside the eigenfrequency range of the h-BN. Note that the color bar for Fig.~\ref{fig_md_strain_0.0}~(c) is three orders smaller than that of Fig.~\ref{fig_md_strain_0.0}~(a).

\begin{figure}[htpb]
  \begin{center}
    \scalebox{1.3}[1.3]{\includegraphics[width=8cm]{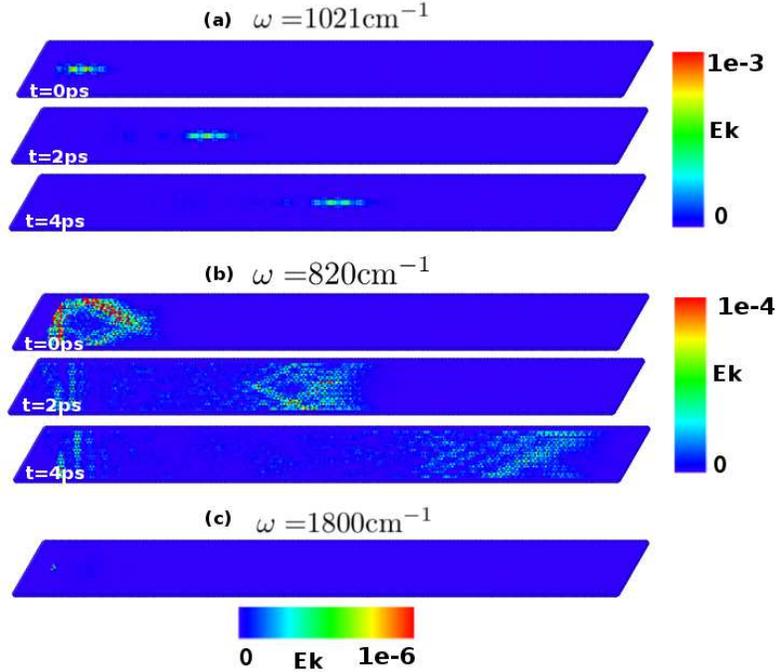}}
  \end{center}
  \caption{(Color online) MD snapshots for the vibrational energy transport along the topologically protected B-B interface in the h-BN ribbon that is biaxially stretched with strain 0.05. The vibrational energy is excited at three representative frequencies, including (a) $\omega=1021$~{cm$^{-1}$} (within the topological branch), (b) $\omega=820$~{cm$^{-1}$}, and (c) $\omega=1800$~{cm$^{-1}$}.}
  \label{fig_md_strain_0.05}
\end{figure}

Fig.~\ref{fig_md_strain_0.05} shows some MD snapshots for the energy transport along the topological interface in the biaxially stretched h-BN with $\epsilon=0.05$. The vibrational energy can be efficiently transported by the topologically protected interface modes ($\omega=1021$~{cm$^{-1}$}), with 94\% transmission. For a normal phonon mode ($\omega=820$~{cm$^{-1}$}), the vibrational energy can be transported less efficiently, with 26\% transmission.  As before, there is almost no energy injection for frequencies outside the eigenfrequency range of the h-BN ($\omega=1800$~{cm$^{-1}$}). \rev{We note that these frequencies are for the biaxially stretched h-BN system.}



\section{Conclusion}

In conclusion, we have studied the modulation of topologically protected phonons in two-dimensional monolayer h-BN through strain engineering, including biaxial, uniaxial, and shear strains. We find that the frequency gap of the topologically protected phonons is rather robust and cannot be closed for all types and magnitudes of mechanical strains.  While the frequency gap cannot be completely switched off, we have demonstrated that the position and magnitude of the topological bandgap can be tuned efficiently through biaxial tensile strains.  More specifically, the position of the frequency can be reduced from 1200~{cm$^{-1}$} to 700~{cm$^{-1}$} by a biaxial strain $\epsilon=0.12$. \rev{As a result, the topological phonon is able to transport the vibrational energy with high transmission even if the topological interface undergoes significant mechanical deformation.} These results may be of practical significance towards the study and development of actively tunable phononic topological materials and structures.

\textbf{Acknowledgements} The work is supported by the Recruitment Program of Global Youth Experts of China, the National Natural Science Foundation of China (NSFC) under Grant No. 11504225, and the Innovation Program of Shanghai Municipal Education Commission under Grant No. 2017-01-07-00-09-E00019. HSP acknowledges the support of the Mechanical Engineering department at Boston University.



%
\end{document}